**Environmental flow for Monsoon Rivers in India:**

**The Yamuna River as a case study**

[1]Vikram Soni, [2]Shashank Shekhar, [3]Diwan Singh

1. Jamia Millia Islamia University, New Delhi.

   Email: vsoni.physics@gmail.com

2. Department of Geology, University of Delhi, Delhi-110007.

   Email: shashankshekhar01@gmail.com

3. Natural Heritage First, B 49 A , Vasant Vihar, New Delhi -110057

   Email: diwans2007@gmail .com

**Abstract**

We consider the flows of Monsoon Rivers in India that will permit the river to perform all its natural functions.  About 80% of the total flow for Indian rivers is during the monsoon and the remaining 20% is during the non monsoon period. By carrying out a case study of the river Yamuna in Delhi we find that at least 50% of the virgin monsoon (July to September) flow is required for the transport of the full spectrum of soil particles in the river sediment. A similar flow is needed for adequate recharge of the floodplain aquifers along river. For the non monsoon period (October to June) about 60% of the virgin flow is necessary to avoid the growth of still water algae and to support river biodiversity.

**Introduction**

The river, its catchments, basin and flood plains is a water course made by nature over an evolutionary timescale of tens of millions of years, which once damaged, often, cannot be fully reclaimed. Rivers in India are now overdrawn, silted and polluted. The health of the river system, then, becomes a priority.

Indian rivers are Monsoon Rivers which have deep and wide floodplain aquifers that run for thousands of kilometers and are an enormous natural storage for water (Soni et al. 2009, 2013). They get recharged by the river flow and floods, especially during the monsoon. They feed groundwater aquifers in their environs. Given the increasing scarcity of water they are likely to be one of few perennial water resources of the future.

 If not polluted by human intervention, the water in the river is the best water. In millions of years of flow the river has washed down salinity and other contaminants to give really good quality 'mineral water', which makes a river a special and vital survival resource. Priority must rest in the sustainability of all the ecological services the river provides. Care has to be taken to avoid injury to the catchments, floodplains and the basin by overbuilding dams, canals and from the discharge of pollution in rivers. For this environmental flow have to be maintained.

**Preserve and Use**

In the present circumstances of water scarcity due to rising populations, agriculture demands and urbanization, the question that is  assuming immense importance is how to use river water resource in a non invasive way – 'Preserve and Use'. *What is the maximum sustainable water use from a river that will maintain the ecological integrity of a river as a perennial resource?*

Like any other living being the river needs to cleanse itself. This happens when the big rain makes the river somewhat fast and furious, flushing out pollution, silt and the still water algae. For Indian rivers this happens during the monsoon, when 80% of all rain falls each year. It is during these periods of heavy rain that the river performs many essential functions. The much higher discharge during the monsoon markedly increases the velocity and depth of flow. This is essential for soil transport to avoid siltation of the river bed – a dire need for our rivers today. The rise in the level of the river makes its reach and breadth larger. Flooding inundates the floodplains and large tracts of surrounding land recharging the ground water. The faster flow also increases the water pressure on the sides which is again good for recharge of the environs. Remember, some of this flow is returned to the river by the floodplain during the lean season. This wetting of the surrounding areas also revives the good bacteria in the subsoil that clean up pollution. At the mouth or delta of the river it performs the very important function of stopping the saline ingress of the sea, supporting the barrier mangrove ecosystems and producing a very fertile interface between the river and the sea both in terms of soil and fish – on which depends the livelihood of millions. We have to look upon the large monsoon flow as an essential. If we cut this flow drastically, many of nature's functions and bounties that we take for granted will be disrupted.

A report by the International Water Management Institute (IWMI 2005) indicates that if a river is to be maintained close to its pristine state approximately 60-80% of its total mean annual flow is required. The actual water use to avoid environmental stress should be not more than 50 % of the total annual flow of the river (Fig.2; IWMI 2005).

**Present River Flows: Yamuna as a Case study**

The first question that arises is the data on river and its flow characteristics. We shall carry out this exercise for the Yamuna in Delhi stretch (Fig.1). There is a problem here. There are many sets of data available on flows and discharges but they do not tally.

These flows must square up with such river parameters as the depth of the river channel during monsoon and non monsoon periods, the width of the river channel and the velocity of the flow. These parameters vary along the river course so we can only give averages. This determines the velocity of the flow – obviously the flow must be faster when the channel is narrower. When the flow is faster the river transports more sand and other particulate matter from the river bed.

At the outset we would like to point out that subject of river flows is not an 'exact' or reductionist science like space science or nano-science. It has to be holistic and convergent and thus approximately sensible. Besides, we have to factor in that rivers are heavily used for agriculture, domestic supply in cities and for industrial supply hence we cannot always demand pristine river flows. We have to safeguard all the essential natural functions of the river and yet be practical. Below, we shall use average parameters to find a consistent picture of the river dynamics.

**River Flow Analysis**

The total mean annual flow for the Yamuna in the Delhi stretch (Fig.1) is about 13.9 Thousand million Cubic Meter (TMCM) (Jha et al. 1988). The water abstraction from river Yamuna for irrigation and drinking water uses till Delhi is about 9.5 TMCM (CPCB 2006). This leaves only a free river flow of about 4.4 TMCM. For the Yamuna, as for most Indian rivers, ~ 80% of the flow (~ 11 Thousand million Cubic

Meter (TMCM)) occurs in the monsoon (July to September) and only 20% (2.8 TMCM) in the rest of 9 non monsoon months (January to June and October to December) (CPCB 2006; Jha et al. 1988).

Data from the Flood and Irrigation department of Delhi, on monthly release of water from the Wazirabad barrage (Fig.1) in Delhi, indicates total release of 3.8 TMCM in the monsoon and 0.3 TMCM in the rest of 9 non monsoon months (Babu et al. 2003) .

Our estimate of the river flow is based on the following average set of data

(i)    The river cross section which includes (a) the main channel of the river (measured at many stretches and times) and averages to 80 meters in width, which is occupied through the year, in particular during the non monsoon period (b) the gently sloping shoulder area outside the main channel, which is occupied by water when the river stage increases beyond 207 meters above mean sea level (Mamsl). This is shown in Fig. 2.

(ii)   The column of water in the main channel averages to about 0.6 meters during the non monsoon period and rises to an average 2.5 meters during the monsoon period. We assume a rectangular cross sections of the main channel (a) of width 80 meters and water column depth 0.6 meters during the non monsoon period (b) of width 80 meters and water column depth 2.5 meters during the monsoon period.

(iii)  During monsoon period the gently sloping shoulders have an average water depth of 1.9 meters at the main channel which goes to zero depth at the outer width. Thus for monsoon flow through shoulder areas adjacent to the main river channel we can approximate this by a rectangular cross section of the mean average water column depth ~ 1 meter and a width of 125 meters on each side of the main river channel.

(iv)   The measured average velocity of the flow during the non monsoon period ~ 0.4 m/s.

(v)    The average slope of water surface is 0.00041.

From this data we determine the hydraulic radius of the flow and by using appropriate roughness coefficients which decrease proportionately with increase in river stage ( see, for example, Hardy et al. 2005 and supplementary) for the flows in the main channel and shoulder during the monsoon and non monsoon periods, we determine the Manning velocities for different stages and width sections and the corresponding discharge. The data was used to prepare a scatter plot of total river discharge (including shoulder flow adjacent to the main river) versus river stage and an appropriate trend line was fitted in the plot. Thus a standard rating curve giving site a specific stage discharge relationship was prepared (Fig.3). A plot of river stage versus main river channel flow velocity for high stage flood including peak flow (Fig.4) was prepared to have an insight in to relationship between river stage and main river channel flow velocity.

In the  non monsoon period, the average river stage of 207 Mamsl corresponds to a river channel discharge of 19 m$^3$/s (Fig.3). As mentioned above the average monsoon river stage is 1.9 meters above the non monsoon rive stage of  207 mamsl, hence the average monsoon river stage in the river channel was estimated as 208.9 mamsl. This corresponds to the main channel river flow velocity of ~1.6 m/s in the monsoon period (Fig.4) and total river  discharge of 507 m$^3$/s (Fig.3). The results are detailed in the supplementary section-1.  We find good correspondence between our calculated discharge and the observed discharge.

**Non monsoon flow**

The total present non monsoon flow in nine months was estimated at 0.44 TMCM (Section-1 of supplementary information). As mentioned above, the total mean annual virgin flow of river Yamuna is 13.9 TMCM and 20% of this, that is 2.8 TMCM, is the non monsoon virgin flow of river Yamuna in Delhi stretch (Fig.1). The present total non monsoon flow in nine months calculated above as 0.44 TMCM is only 16% of the non monsoon virgin flow of river Yamuna. This is not the ideal non monsoon flow required for a river to perform its various functions.

**Monsoon flow**

The river flow regime widens considerably in the monsoon months and the flow regime can be segregated into river flow through the main river channel and through areas adjacent to the main river channel (Fig.2). The total monsoon flow in main river channel in three months was estimated at 2.5 TMCM and in the shoulder areas adjacent to the main channel of river Yamuna it was approximately estimated as 1. 5 TMCM (Table-1 and section-1 of supplementary information).

Thus the present actual monsoon flow of river Yamuna upstream of Wazirabad barrage (Fig.1) was estimated at approximately 4 TMCM, which gives an average monthly flow during monsoon months of approximately 1.3 TMCM/month.

A summary of river flow during different seasons is given as Table-1. It is thus estimated that the average present mean annual flow through this stretch of river Yamuna is 4.44 TMCM. It is approximately 32% of the total mean annual virgin flow of river Yamuna through this stretch. It is to be noted that these discharge figures are average values and they are subject to annual variations.

**Multiple flow requirements**

The question of what constitutes a flushing flow and its significance is a gray zone with varied opinions. Himalayan rivers carry an enormous sand, gravel and silt load in the monsoon. But the increasing diversion of water does not allow for natural flooding. Earlier, flooding used to deposit sand in the floodplains and carry the rest to the sea.  But now, much soil is retained in the river bed.

Thus, one important function of a Himalayan river in the monsoon is to be able flush out this debris to avoid silting of the channel. This can only happen if the force or velocity of the flow is large enough to transport this material all the way to its mouth or delta. This has not been happening in most of the rivers where siltation of the river bed is endemic and has recently also caused unnecessary flooding. Add to this the heavy sewage discharge of grit and even metallic waste into the river. It thus requires a strong flow to carry away these deposits and keep the integrity of the river channel. There is a hierarchy of flows  required for a river to perform its several natural functions. They are discussed separately below:

## 1. Algal Choking

To avoid still water algal growth we need a minimum flow velocity of 0.75 m/s (Chow 1973). The non monsoon flow (9 months) to enable the river Yamuna to do this is estimated to be 1.8 TMCM (section-2 of supplementary information), whereas the present total non monsoon flow is just 0.44 TMCM. The estimated desired minimum non monsoon flow is approximately 60 % of estimated non monsoon virgin flow of 2.8 TMCM in river Yamuna.

## 2. Sediment Transport

According to Brahm's (Airy's) empirical law the rule of thumb is that the relationship (see Wikipedia, Garde 1995 ) between the mass of the particle, m, and the river flow velocity, v,  required to transport it is as follows:

$(m_1/m_2) = (v_1/v_2)^6$

where $v_1$, is the velocity required to transport a particle of mass, $m_1$, and $v_2$ , the velocity required to transport a particle of mass, $m_2$.

Conventionally the particle size is denoted by the diameter and thus  the particle sizes refer to the diameter of the particle. The soil characteristics of the Yamuna flood plain in Delhi (Fig.1) have a large spectrum of particle sizes (diameter) from 30 µm - 1 mm (Babu et al. 2003). For constant density, the mass is approximately proportional to third power of the diameter and thus the size of the transported particles is proportional to square of the velocity (section-3 of supplementary information).

A flow velocity of 0.75 m/s can move only silt size particles of size up to ~ 60 µm (Pettijohn 1984; Chow 1973). Using this as initial data the relationship between the velocity required to dislodge a given size of particle was mathematically extrapolated using Brahm's (Airy's) law to produce a plot showing water flow velocity required to dislodge the different particle sizes (Fig.5). The relationship between flow velocity in the main river channel and total monsoon discharge through river Yamuna is given in Fig.6.

The total monsoon discharge for a set of average monsoon river stages and the corresponding water flow velocity in the main channel and in the shoulder areas is estimated in (Table-2). The average monsoon river stage of 208.9 mamsl corresponds to the discharge of 3.9 TMCM (~36% of the total mean annual monsoon flow) through the river Yamuna (Table-2). The velocity in the main river channel at this total discharge was estimated at ~1.57 m/s (Fig.6; Table-2). This velocity of flow can dislodge sediment particles of the size ~ 265 µm (Fig.5).

The river channel is the faster flowing section of the river and thus will have a larger fraction of the larger soil particles and grit from sewage discharge. An average monsoon discharge of 5.5 TMCM (50% of the total mean annual monsoon flow) through river Yamuna corresponds to a river stage of 209.2 Mamsl (Table-2). The velocity in the main river channel at this total discharge was estimated at 1.8 m/s (Fig.6; Table-2). This velocity of flow can dislodge sediment particles of the size ~ 340 µm (medium sand; Pettijohn 1984) (Fig.5). The main channel can thus be cleared of the larger and heavier sediments more efficiently at this flow velocity. Field based observations find average monsoon peak flow to be approximately 1.4 m above the average monsoon river stage (Table-3). Based on  this  the peak flow associated with the average monsoon river stage of 209.2 Mamsl (corresponding to total discharge of 5.5 TMCM) would be 210.6 Mamsl. The flow velocity in the main river channel at the river stage of 210.6 Mamsl would be 3.4 m/s (Fig.4). This river flow velocity is sufficient to dislodge particles of size ~1200 µm (1.2 mm) (coarse sand; Pettijohn 1984) (Fig.5). Hence this river flow velocity will efficiently remove most of the heavy sediments and grit from main river channel and prevent siltation and shallowing of the river channel.

The water flow velocity in areas adjacent to the main river channel corresponding to this monsoon discharge of 5.5 TMCM at a river stage of 209.2 Mamsl was estimated at 0.93 m/s (Table-2). This flow velocity can dislodge sediments particle of size ~ 95 µm (very fine sand; Pettijohn 1984) (Fig.5). Such well sorted sediments will have higher permeability leading to enhancement in river bank storage during the monsoon floods.

Empirically, it is clear that the river is heavily silted and at present has a depth of only 0.6 meters in the summer! To remove all river bed particles of a diameter up to 1-2 mm (coarse sand to very fine gravel; Pettijohn 1984) will require a monsoon flow larger than 50% ( 5.5 MCM), but we also have to balance this with practical reality ( agricultural needs) . A 50% (5.5 MCM) monsoon flow can take out particles of diameter up to ~ 1.2 mm; however, when such particles are transported and desilting occurs, the main channel will deepen enhancing the flow velocity. Consequently, even particles of larger size will be transported. We conclude that at least 50% (5.5 TMCM) of the monsoon virgin flow of the river Yamuna is the flushing flow required in this stretch.

## 3. The Delta and the sea interface

The fertility of the delta and the sea interface depends crucially on the transport of soil material down to the river mouth, which in turn depends on having reasonable flushing flows. Strong flows are needed to avoid ingress of the sea, preserve the mangrove interface. Thus, only if the flushing flow in the monsoon months is maintained can the environmental requirements of the delta and sea interface region be reasonably met.

## 4. Dolphins and Biodiversity

The river dolphin has disappeared in the Yamuna. The story for fish is similar. The flow prior to Delhi (Fig.1) has a water depth of about 0.6 meters through the peak summer months which is too meager to support a healthy fish population. The flow downstream of Delhi is mainly sewage, for this period, which is again not optimal. Besides, there is a variety of floodplain biodiversity which needs this minimum flow. If the mean annual environmental flow of river Yamuna is maintained at approximately 50-60% of the total annual flow in river Yamuna, it may suffice for the ecological requirement of the river.

## 5. Recharge Flows

The alluvial sandy soil in the Yamuna flood plain has aquifer parameters conducive for recharging the aquifer during a normal monsoon season (Rao et al. 2007). The water flowing through the river Yamuna, the groundwater in river bank storage ( flood plain just by the side of the river) and the other water bodies are in dynamic equilibrium (Shekhar and Prasad 2009). It is thus important to assess the desirable monsoon flow in river Yamuna required for proper recharging of flood plain aquifers.

The groundwater velocity through the sand in the Yamuna floodplain, at peak flood (River stage = 210.3 Mamsl ) has been estimated as 2.12m/day. With this velocity, the *lateral groundwater recharge beyond the flooded stretch* over a period of three months (post peak flood), would be  about 200 meters on either side (Section-4 of supplementary information).

During the peak monsoon flow the river expands to a breadth of ~500 m and recharges the flood plain by  gravity recharge. An additional ~400 m is recharged by lateral groundwater flow . Therefore, the total width of the flood plain recharged by the river during the monsoon as ~ 1 K

It is still not enough to recharge the whole floodplain which stretches at least 5 km. A flow of 50% of virgin flow will certainly widen the reach of the river enabling  more recharge.

Besides, the wetting of the soil is also important as the good bacteria that can breakdown nitrates function only in the presence of water. These are strong reasons to maintain such ecological flows to preserve the recharge and quality of the water provided by the river, as aquifers are getting significantly impoverished

## 6. Dilution Flow

The city of Delhi is an urban mess, with a population touching 20 million which is way beyond its carrying capacity (Soni 2003). Downstream of Wazirabad barrage (Fig.1) a series of about 20 drains flow into the Yamuna making it a giant drain (Shekhar and Sarkar 2013). The city generates ~3.27 Million Cubic Meter (MCM)/day of treated and untreated sewage (CPCB 2006). The annual average Bio-chemical Oxygen Demand (BOD) varies from 11 mg/l to 24 mg/l in the river water. The desired standard for BOD level (CPCB 2006) in the river water is 3 mg/l. Thus the average non monsoon season BOD level is close to eight times more than the desirable level. This would require a dilution of 8 times to make it fall into safe health limits - or a fresh water flow of over 0.6 TMCM a month. Clearly, during the monsoon months there is adequate flow for the dilution required to maintain health standards. But during the nine non monsoon months a dilution flow of about 5 TMCM will be required, which is not possible. Thus, in the non monsoon months, there is no option but to treat the sewage of Delhi and reduce the BOD to no more than 5 mg/l. The Yamuna needs a comprehensive cleansing of sewage to be safe.

## 7. Minimum flow

A Memorandum of Understanding (MOU) between basin states of river Yamuna for sharing of Yamuna river water sets aside a minimum flow of 10 $m^3$/s throughout the year, for ecological purposes (NIH 2011). This amounts to a minimum fresh water flow in the river of 0.86 MCM/day, whereas we found that we need about 6.6 MCM/day just to avoid algal choking.

## Conclusion

We have attempted a detailed analysis to work out a genuine ecological flow that can maintain the health of the river and all its natural functions. Our finding is that the river needs close to 50-60 % of the total flow to be safeguarded as free flow, regardless of the season. Briefly, during the monsoon, 50% of the free flow is needed for efficient soil transport (to avoid silting the river bed) and during the non monsoon period, 60% of the free flow is needed to avoid algal choking. Since rivers in India are monsoon Rivers, we feel that setting a single norm of 50-60% of the total flow as a free flow, all year round, is appropriate for all rivers.

The long term consequences of overexploiting the river and cutting flows will terminally affect the river and its surrounds. In this context the river Yamuna is already overexploited (Khosla and Soni 2012) (section-5 of supplementary information). Our suggestion is that since the lean season use is mainly for agriculture, restoration of flow in water short rivers can be accomplished by water harvesting, more efficient agricultural practices like drip irrigation and moving more water intensive agriculture ( for example for rice and sugarcane ) to water surplus areas.

## Acknowledgements

The cooperation of Shri Vikas Jayani is duly acknowledged for helping in verifying the river channel data using remote sensing imagery. The support and cooperation of Dr. S.V.N Rao, WAPCOS, Hyderabad is duly acknowledged.

**Figure Legend**

Fig.1. Location map of River Yamuna.

Fig.2. A generalized river cross section from Palla area, north of Wazirabad (see Fig.1).

Fig.3. The stage – discharge rating curve for the site near Palla upstream of Wazirabad (see Fig.1) in Delhi stretch of river Yamuna.

Fig.4.  A plot of river stage versus main river channel flow velocity.

Fig.5. A plot of particle size versus velocity of flowing water required to dislodge the particle.

Fig.6. A plot of total monsoon discharge through main channel of river Yamuna and adjacent areas in Palla area of Delhi versus flow velocity in Main River channel.

**Table Legend**

Table-1. A summary of river flow during different seasons.

Table-2. Average monsoon flow statistics through river Yamuna at Palla corresponding to different river stages.

Table-3. Difference between average monsoon river Yamuna stage and peak flow river stage at Palla, Delhi (Source: Irrigation and Flood Control Department, Delhi)

# Tables

| Season | River flow (TMCM) | | | Average river stage (Mamsl) |
|---|---|---|---|---|
| | **Main channel** | **Shoulder flow** | **Total** | |
| **Total Monsoon** (Three months) | 2.5 | 1.5 | 4.0 | 208.9 |
| **Total Non monsoon (Nine months)** | 0.44 | 0 | 0.44 | 207 |
| **Total** | 2.94 | 1.5 | 4.44 | -- |

**Table-1: A summary of river flow during different seasons.**

| Main River Yamuna channel of width 80 m and particle size refers to diameter of the particles | | | | | Shoulder flow area adjacent to the main channel | | | | Total | | Total flow as % of |
|---|---|---|---|---|---|---|---|---|---|---|---|
| River Stage (Mamsl) | Total column of water (m) | Velocity (m/s) | Particle size transported (micro meter) | Discharge (m³/s) | Average column of water (m) | Flow width (m) | velocity (m/s) | Discharge (m³/s) | Discharge (m³/s) | Monsoon discharge TMCM) | Virgin monsoon flow |
| 208. | 2.5 | 1.5 | 265 | 316 | 0.9 | 25 | 0.7 | 191 | 507 | 3.9 | 36 |
| 209 | 2.6 | 1.6 | 290 | 344 | 1.0 | 26 | 0.8 | 223 | 567 | 4.4 | 40 |
| 209. | 2.7 | 1.7 | 310 | 374 | 1.0 | 27 | 0.8 | 259 | 633 | 4.9 | 45 |
| 209. | 2.8 | 1.8 | 340 | 405 | 1.1 | 28 | 0.9 | 299 | 704 | 5.5 | 50 |
| 209. | 2.9 | 1.8 | 370 | 439 | 1.1 | 30 | 0.9 | 343 | 782 | 6.1 | 55 |
| 209. | 3.0 | 1.9 | 415 | 475 | 1.2 | 31 | 1.0 | 393 | 868 | 6.7 | 61 |

Table-2 : Average monsoon flow through river Yamuna at Palla at different river stages

| | Difference between average monsoon river stage and peak flow river stage (m) |
|---|---|

| Year | July | August | September | Average |
|------|------|--------|-----------|---------|
| 2008 | 1.1 | 1.1 | 2.1 | 1.4 |
| 2009 | 0.9 | 1.5 | 0.8 | 1.1 |
| 2010 | 2.1 | 1.6 | 1.3 | 1.7 |
| Average difference between monsoon and peak flow river stage (m) | | | | 1.4 |

Table-3

# Figures

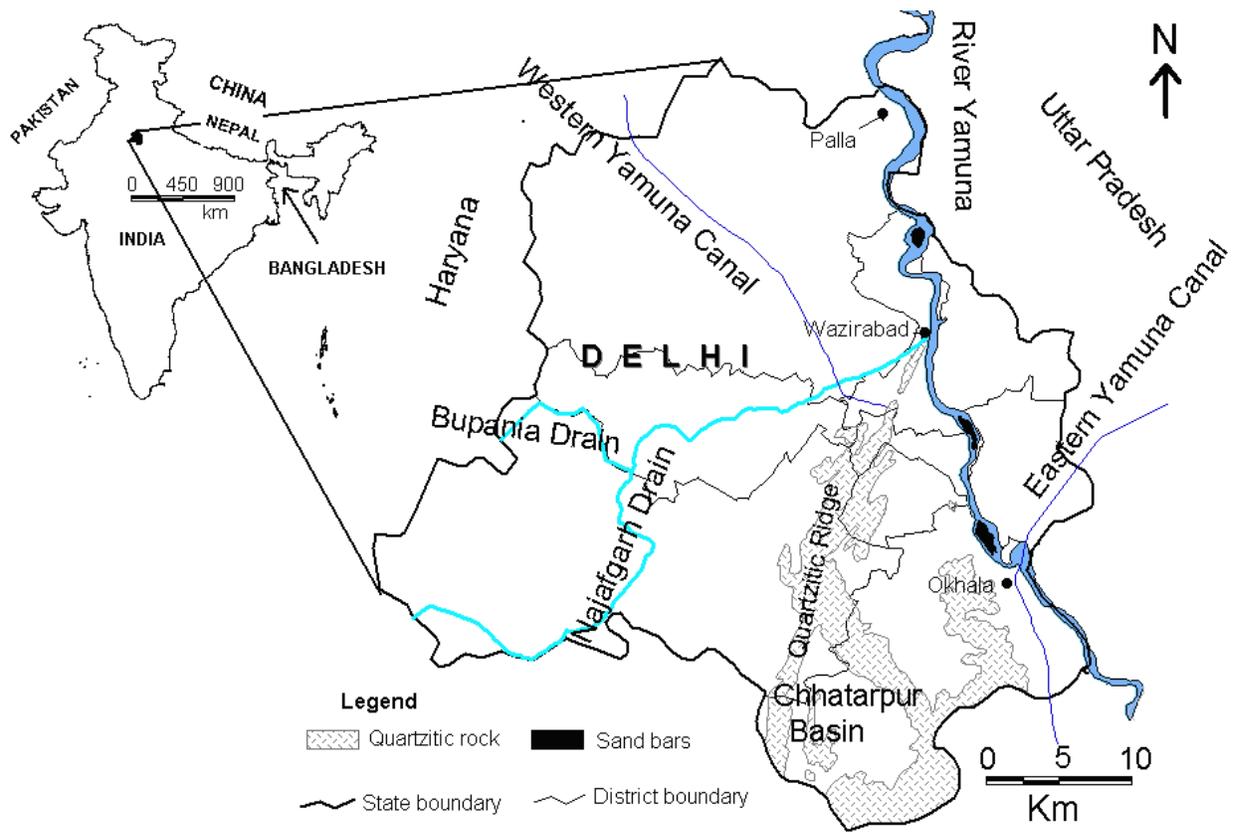

Fig.1

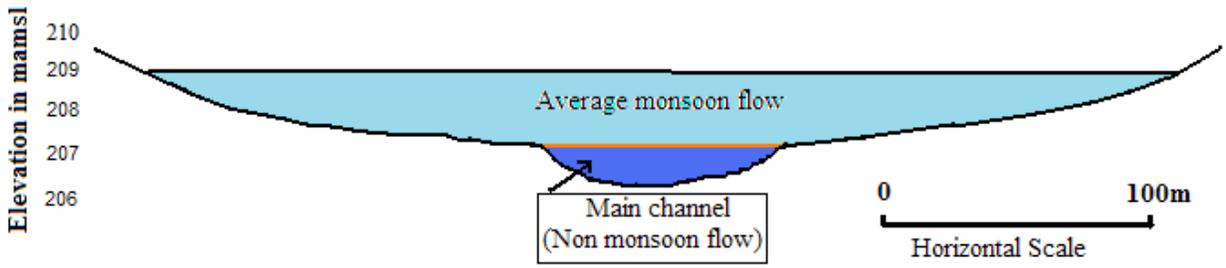

Fig. 2

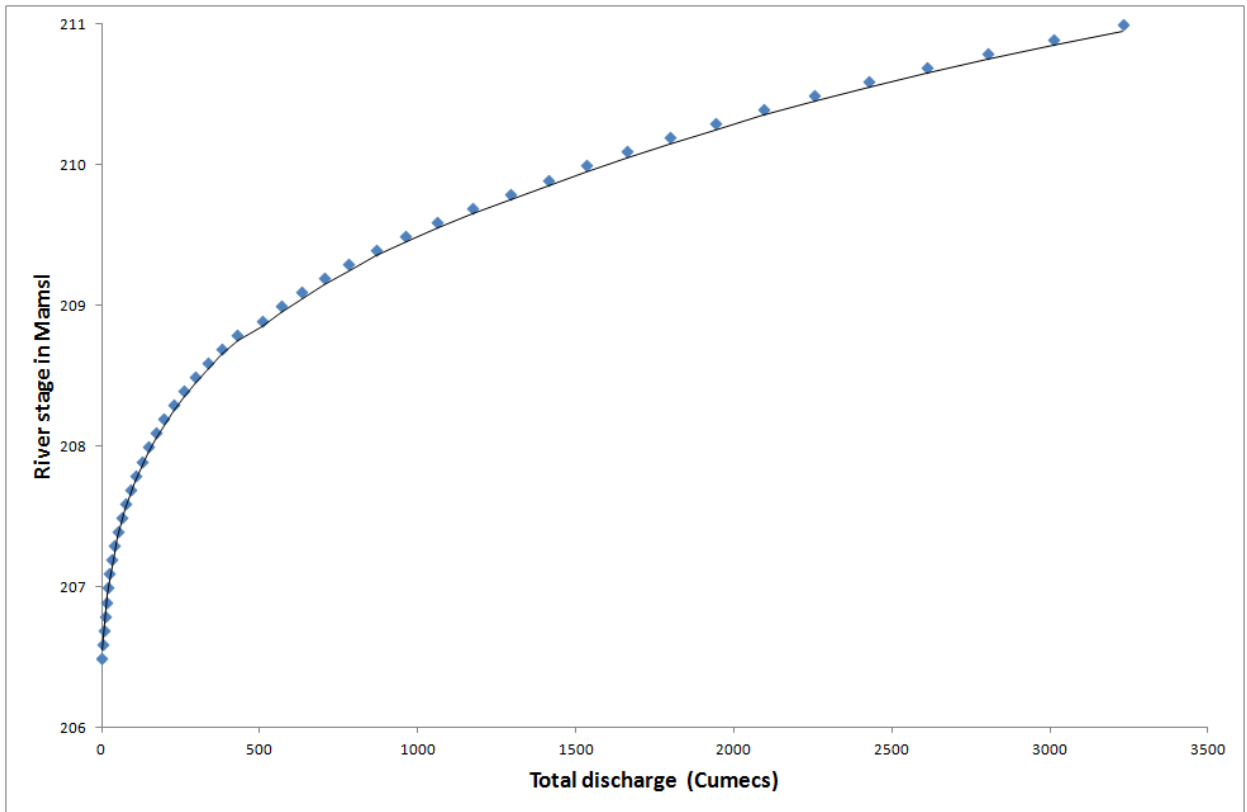

Fig. 3

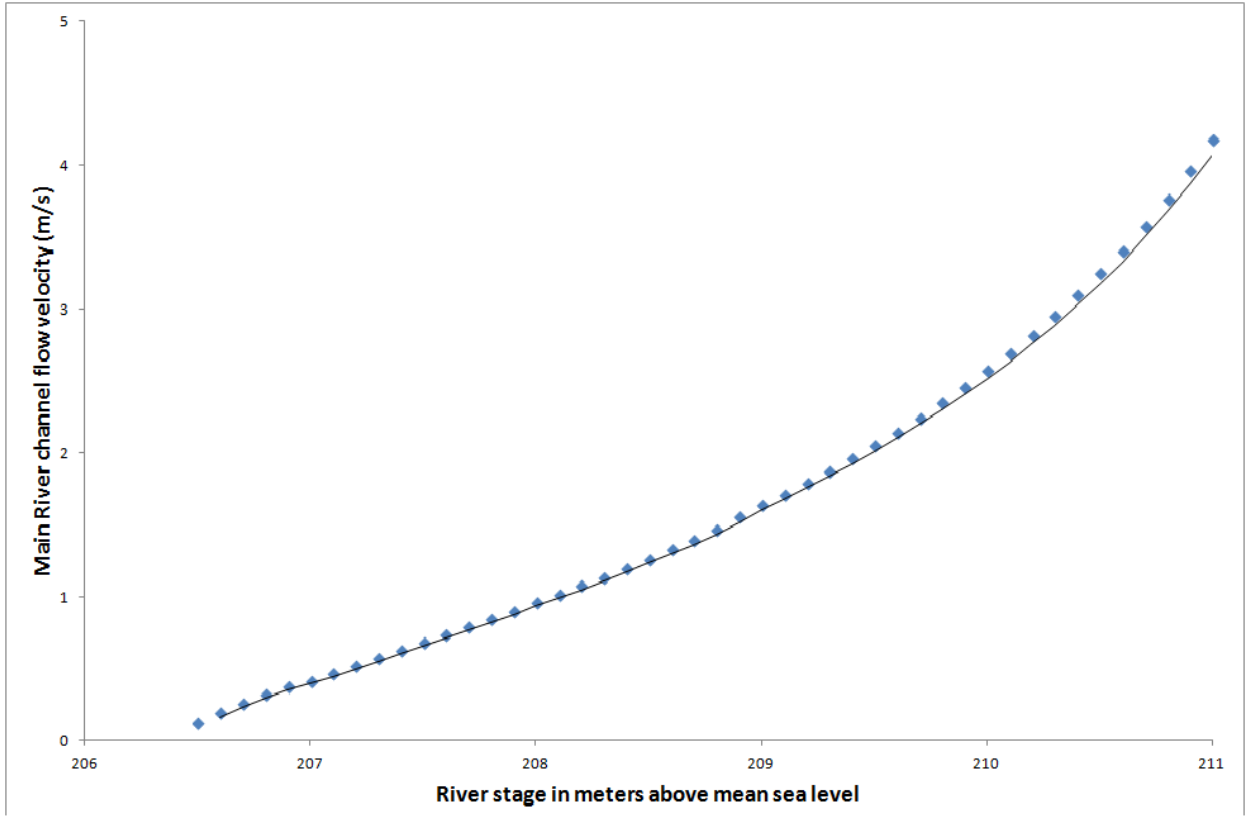

Fig.4

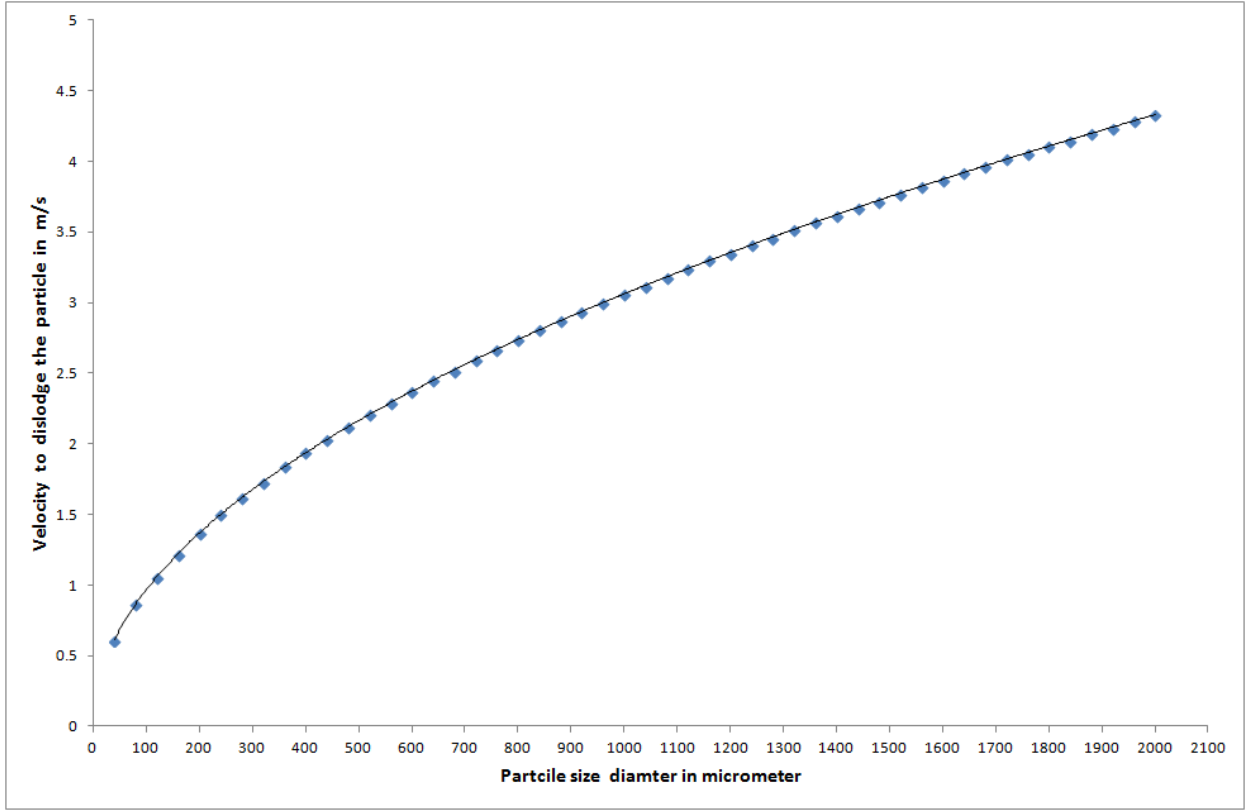

Fig. 5

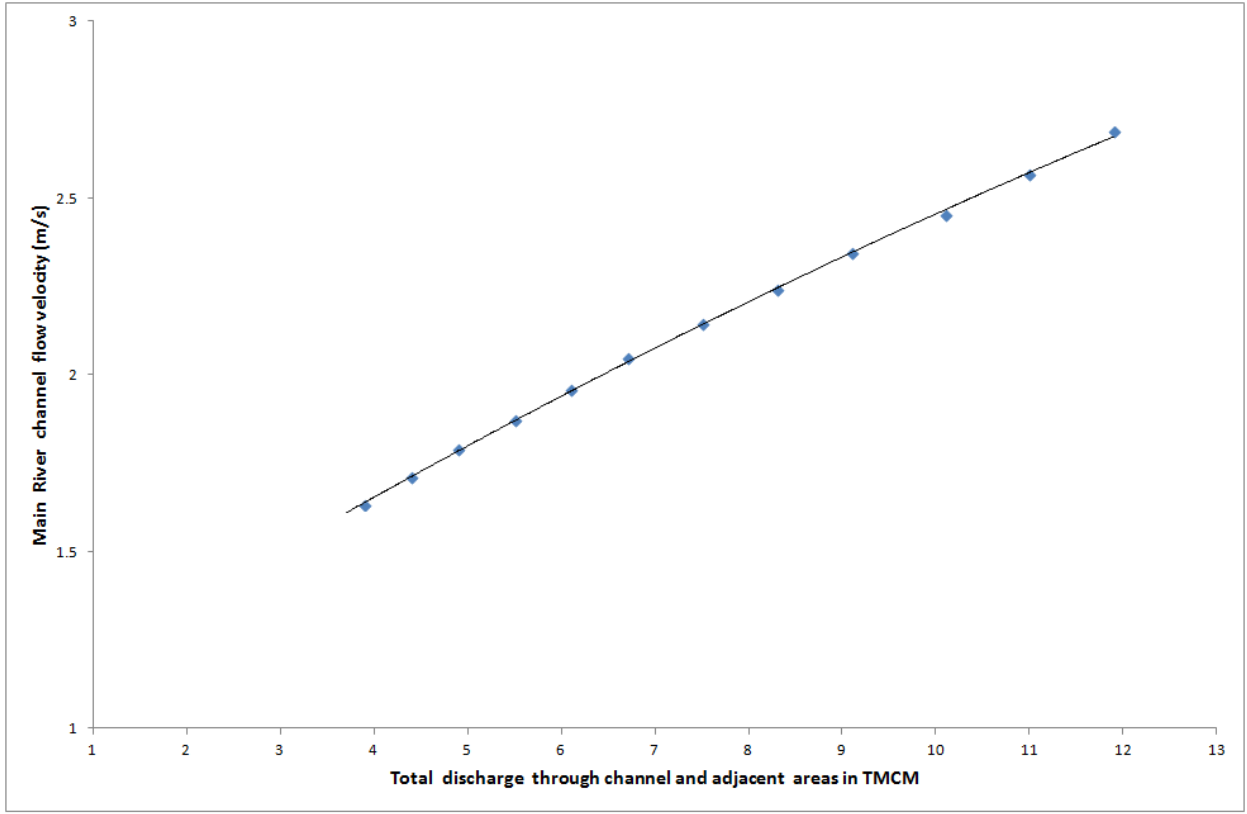

Fig. 6